\documentclass[aps,prl,reprint,superscriptaddress,showpacs]{revtex4-1}

\usepackage{graphicx}
\usepackage{color}
\usepackage{booktabs}
\usepackage{bm}
\usepackage{longtable}
\usepackage{amsmath}
\usepackage{dcolumn}
\usepackage{placeins}
\usepackage{expdlist}

\begin{document}

\title {Spin noise spectroscopy of a noise-squeezed atomic state }

\author{V. Guarrera} 
\email{v.guarrera@bham.ac.uk}
\affiliation{Midlands Ultracold Atom Research Centre, School of Physics and Astronomy, University of Birmingham,
Edgbaston, Birmingham B15 2TT, United Kingdom}
\author{R. Gartman}
\affiliation{National Physical Laboratory, Hampton Road, Teddington, TW11 0LW, United Kingdom}
\author{G. Bevilacqua}
\affiliation{DIISM, Universit\`a di Siena, via Roma 56, 53100 Siena, Italy}
\author{W. Chalupczak}
\affiliation{National Physical Laboratory, Hampton Road, Teddington, TW11 0LW, United Kingdom}

\date{\today}

\begin{abstract}
Spin noise spectroscopy is emerging as a powerful technique for studying the dynamics of various spin systems also beyond their thermal equilibrium and linear response. Here, we study spin fluctuations of room-temperature neutral atoms in a Bell-Bloom type magnetometer. Driven by indirect pumping and undergoing a parametric excitation, this system is known to produce noise-squeezing. Our measurements not only reveal a strong asymmetry in the noise distribution of the atomic signal quadratures at the magnetic resonance, but also provide insight into the mechanism behind its generation and evolution. In particular, a structure in the spectrum is identified which allows to investigate the main dependencies and the characteristic timescales of the noise process. The results obtained are compatible with parametrically induced noise squeezing. Notably, the noise spectrum provides information on the spin dynamics even in regimes where the macroscopic atomic coherence is lost, effectively enhancing the sensitivity of the measurements. Our work promotes spin noise spectroscopy as a versatile technique for the study of noise squeezing in a wide range of spin based magnetic sensors.

 \end{abstract}

\maketitle

\textit{Introduction.} In standard spin noise spectroscopy (SNS), spontaneous fluctuations of the atomic spins in thermodynamic equilibrium can provide information on the system properties, such as resonance frequencies and rates of the decoherence processes \cite{Crooker2004, Mihaila2006, Katsoprinakis2007, Chalupczak2011, Zapasskii2013, Sinitsyn2016}. The extraction of information from the noise in the system is usually based on the fluctuation-dissipation theorem, which relates the spontaneous spin fluctuations spectrum to the linear response of the atomic system to perturbations \cite{Callen1951}. Thus, standard noise studies are specifically done in unperturbed systems, with noise level readout performed via Faraday-rotation type measurements, enabling non-demolition observation of the spin dynamics \cite{Takahashi1999}. In recent years, first attempts to explore SNS beyond the regimes of thermodynamic equilibrium and linear response have appeared \cite{Glasenapp2014,Poshakinskij2020,Swar2018}. This is of interest as, for example, standard spin noise spectra cannot typically disclose information on the system's response to resonant driving fields (out-of-equilibrium dynamics), or reveal the (linear and non-linear) couplings between the spin coherences associated with the system's relevant energy levels \cite{Glasenapp2014}.  

In the present work, we employ SNS to experimentally investigate the noise spectrum of an atomic spin system exposed to a Bell-Bloom type excitation \cite{Bell1961, Gartman2015}. This system undergoes indirect pumping, via optical excitation and spin-exchange collisions (SEC), and thus a non-equilibrium ensemble polarization is created. In addition, it can be parametrically excited by proper modulation of the optical pump amplitude, as we have previously demonstrated in \cite{Guarrera2019}. Standard SNS is often a clever diagnostic technique \textit{alternative} to a pump/probe configuration, however we demonstrate that, in the presence of an active pumping mechanism, SNS can also provide useful, complementary information on the system dynamics. In particular, it can assist in identifying the mechanism leading to the formation of noise squeezing, which we previously observed in the transient dynamics of conjugated spin variables following the parametric excitation \cite{Guarrera2019}. As such, on one side our work extends the scope of SNS beyond its current boundaries, which has long been appealing from both a theoretical and experimental point of view \cite{Li2013,Glazov2013}. On the other side, it promotes SNS as a powerful technique for studying spin squeezing. Noise squeezing has attracted much interest in recent years as a strategy to effectively improve sensors' performances beyond those set by classical, and standard quantum limits. It has been achieved in several out-of-equilibrium, non-linear systems, where, so far, spin fluctuations have mainly been characterized for the quantification of the amount of squeezing achieved (for a review see \cite{Pezze2018}).



\begin{figure*}
\centering
\includegraphics[width=\columnwidth]{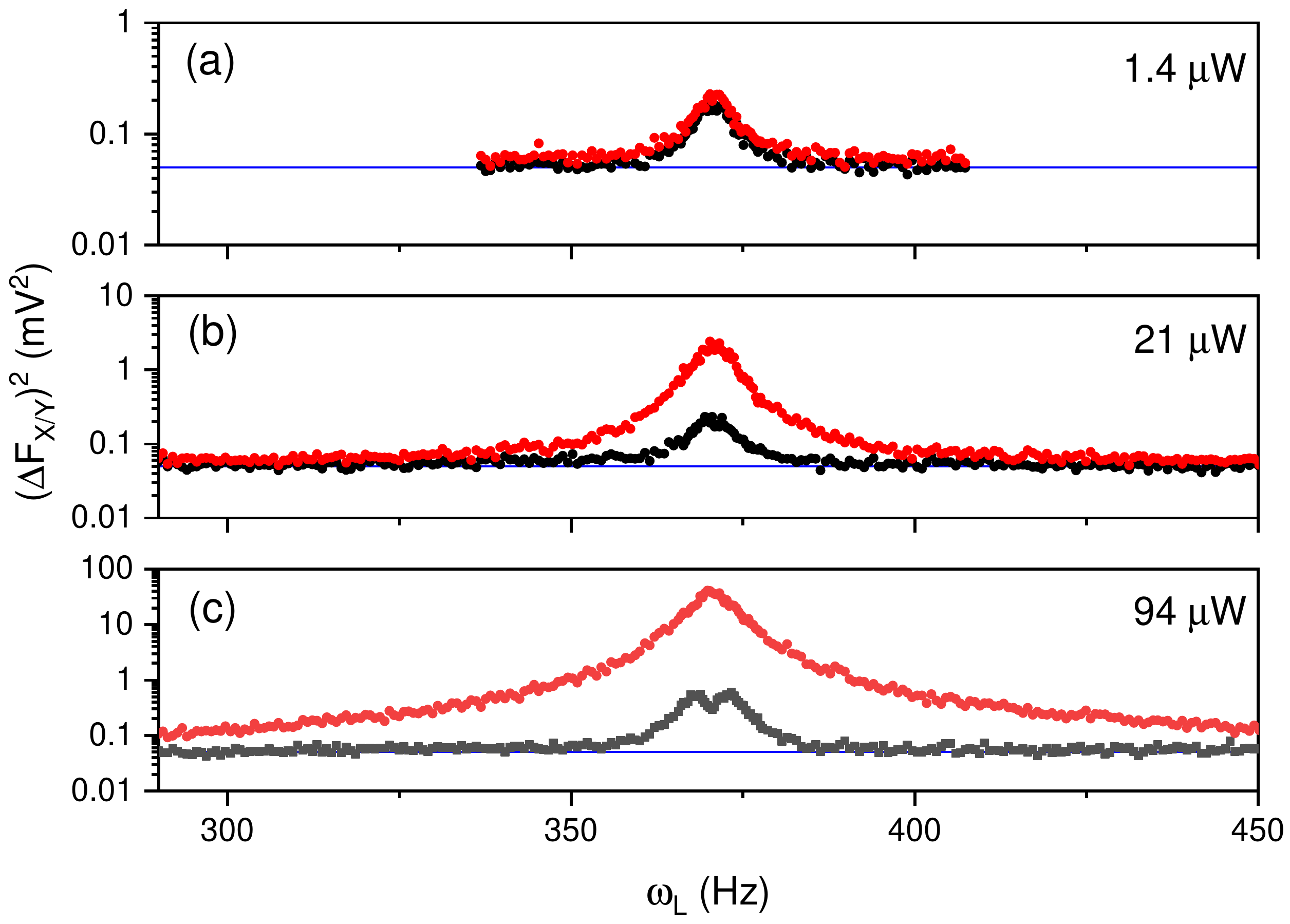}
\includegraphics[width=0.43\textwidth]{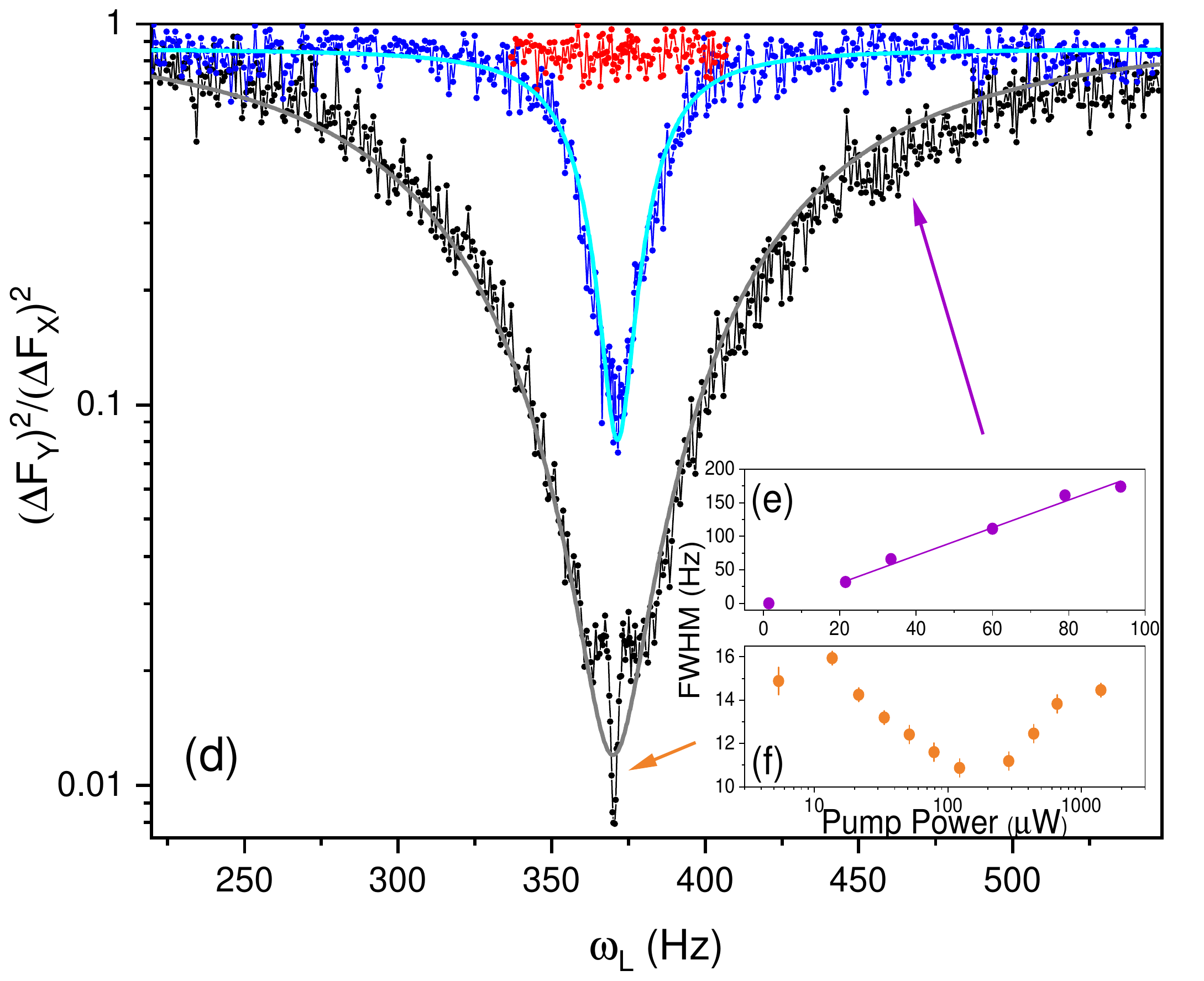}
\caption{ (color online) Dependence of the signal variance, $(\Delta F)^2$, on the Larmor frequency recorded with pump beam powers of (a) 1.4 $\mu$W, (b) 21.5 $\mu$W, and (c) 94 $\mu$W. d) Dependence of the ratio of the two noise components, $(\Delta F_Y)^2/(\Delta F_X)^2$, on the Larmor frequency recorded with pump beam powers of 1.4 $\mu$W (red line), 21.5 $\mu$W (blue line), and 94 $\mu$W (black line). Grey/light blue solid lines in (d) are Lorentzian fits to the tails. Insets: (e) FWHM of the broad feature, and (f) FWHM of the narrow near-resonance feature as a function of pump power. The signal is acquired with lock-in reference to $\omega_{ref}=370$ Hz, and time constant of $\tau=100$ ms (e) and $\tau=3$ ms (f), the latter corresponding to the best time resolution of our lock-in amplifier. The optical excitation modulation frequency is equal to $2 \omega_{ref}$. All the measurements have been done with probe beam power of 146 $\mu$W. At this power, the probe beam does not have any significant effect on the observed noise spectra.}
\label{fig:Spectrum}
\end{figure*}

\textit{Experimental system.} We detect the amplitude fluctuations in the collective spin of an atomic Cs vapour precessing around a static magnetic field at the Larmor frequecy $\omega_L$. Measurements are performed on a room-temperature vapour housed in a paraffin coated glass cell, with an atomic density of $0.33 \times10^{11} \text{cm}^{-3}$. The cell is surrounded by five layers of $2$ mm thick mu metal shields to reduce static magnetic background. A small static magnetic field is created by a Hemholtz pair of coils (along the z-axis, see \cite{Guarrera2019} for further experimental details). The optical pumping is performed by a circularly polarised laser beam frequency locked to the caesium $6\,^2$S$_{1/2}$ F=3$\rightarrow{}6\,^2$P$_{3/2}$ F'=2 transition (propagating along the x-axis). We modulate the amplitude of the pump laser power with a pulse duty cycle of $7\, \%$. The signal produced by the $F=4$ ground state atomic coherences is read out by a probe beam propagating in a direction orthogonal to the pump beam, whose frequency is blue-detuned by roughly 1 GHz with respect to the $6\,^2$S$_{1/2}$ F = 4$\rightarrow{}6\,^2$P$_{3/2}$ F' = 5 transition frequency. The probe light transmitted through the cell (along the y-axis) is analysed by a polarimeter and processed by a lock-in amplifier. 
In this system, we have previously demonstrated that signal amplification at $\omega_L$ and noise squeezing are generated at parametric excitation resonances, i.e. for modulation frequencies of the pump beam amplitude which are equal to $2\omega_L/n$, being $n$ an integer number. In the model we discussed in \cite{Guarrera2019}, the parametric term enters the individual atoms' spin evolution via a sufficiently strong pump, and its macroscopic effect relies on the synchronised atomic response \cite{Fernholz2008}. Also, we recall that in the case of modulation frequency $\omega_M\neq \omega_L/n$, the spectrum of the evolving ground state coherences contains mainly two components: a steady state oscillation at the modulation frequency $\omega_M$, and a transient oscillating at $\omega_L$ damped on a timescale associated with the rate of spin-exchange collisions. 
While our model of Ref.\cite{Guarrera2019} explains well the main experimental findings on the transient oscillation, it provides also an indication that the overall spin dynamics might be complicated by additional non-linear effects, such as those described in \cite{Gartman2018}, and by the role of off-resonant pumping, which might become manifest in the steady state regime \cite{note1}. Thus, it appears especially desirable for such a complex system to rely also on a complementary characterization of the spin coherences, and to access them even when their collective signal is lost (e.g. in steady state).

\textit{Spin noise spectrum.} In the following we focus our analysis on the case with modulation frequency of the pump close to $2 \omega_L$. This particular parametric resonance has been chosen due to the absence of any direct driving contribution to the coherent signal at $\omega_L$, i.e. $\omega_M \neq \omega_L/n$. Also, the sensitivity of the lock-in amplifier in this case can be set to high-level, which enables noise studies with high dynamical range across a wide range of pump beam powers. The signal created in the experiment is acquired for up to 30 seconds, lock-in referenced to $\omega_{ref}=\omega_M/2=370$ Hz, and recorded in its in-phase/out-of-phase (X/Y) components. This means that the detected signal takes the form $S(t)=F_X(t) \cos(\omega_{ref} t)+F_Y(t) \sin(\omega_{ref} t)$ in the reference set by the lock-in amplifier, where $S(t)$ is proportional to the collective spin component along the probe beam direction.
The pump beam amplitude modulation frequency is fixed while the strength of the static magnetic field, and thus $\omega_L$, is scanned across the resonance $\omega_{ref}=370$ Hz. An acquisition board records several hundred of samples of the lock-in outputs and subsequently the variances of the X and Y lock-in signal components are calculated at a time T from the starting of the optical pumping. We note that, when performing the SNS measurements well into the steady state regime (which is also instrumental for ideal lock-in resolution), the coherent signal of the collective atomic spin is mainly lost \cite{note1}, as if the atomic system was effectively unpolarized. 

However, the noise spectrum already shows a clear resonant signal for relatively low pump powers, with similar values of the X and Y variances, see Figure \ref{fig:Spectrum}(a). Note that with our setup, the spectrum is obtained by varying together $(\omega_L-\omega_M)$ and $(\omega_L-\omega_{ref})=(\omega_L-\omega_{M}/2)$, as this allows us to cancel the intrinsic sensitivity of the lock-in amplifier to phase noise of the reference signal and to \textit{critically} improve the accuracy of the noise detection. The obtained spectra allow us to easily single out the atomic contribution from the total noise. Indeed, driving the Larmor frequency away from the lock-in reference allows to access the background noise (marked with a solid blue line in Figure \ref{fig:Spectrum}(a)-(c)) mainly due to photonic shot-noise, as we have verified by measuring its linear dependence on the probe beam power. 

\begin{figure}[h!]
\includegraphics[width=\columnwidth]{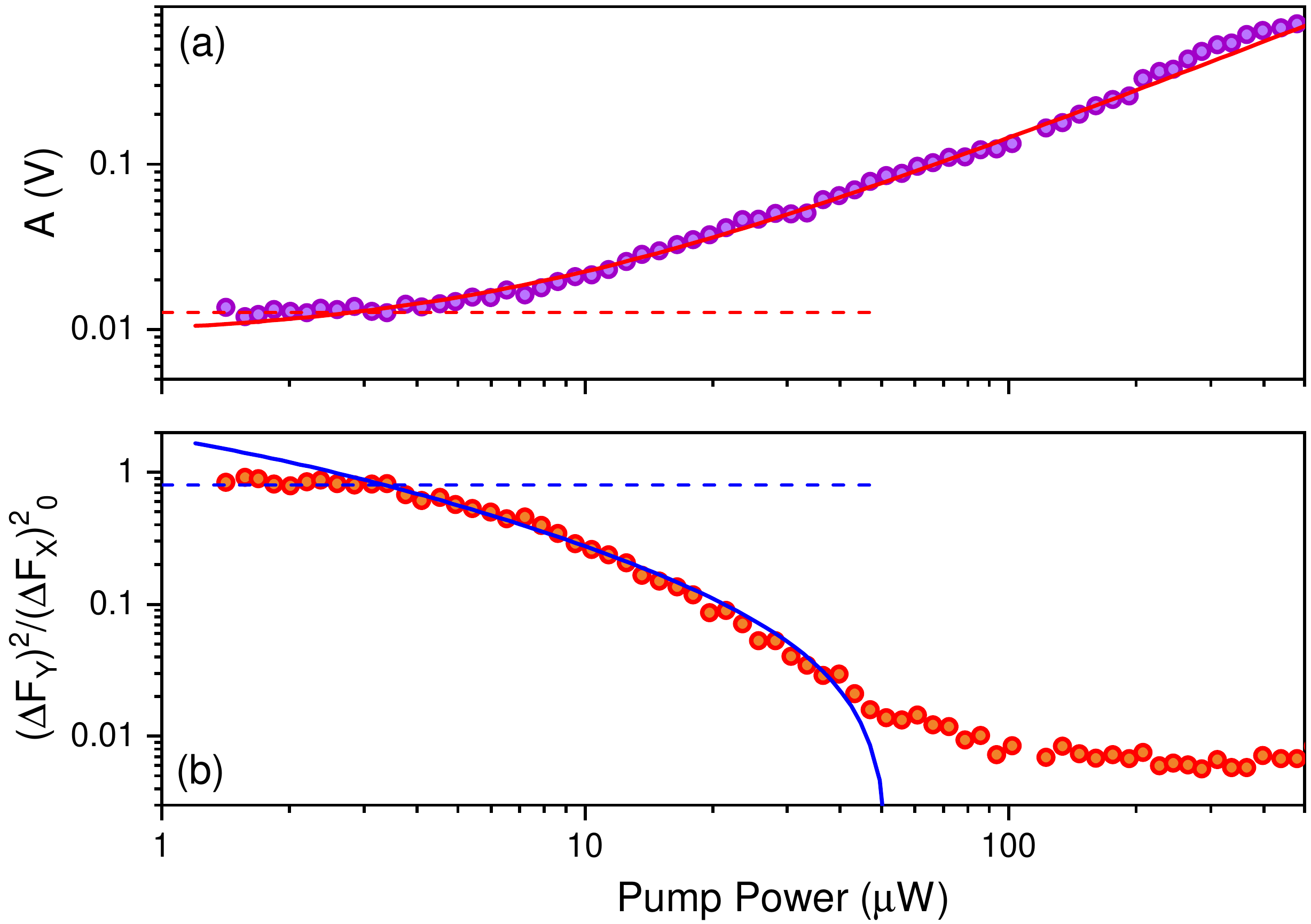}
\caption{ (color online)(a) Integrated noise signal as a function of the pump beam power. Solid line is a fit with a linear function, dashed line is a fit of the initial plateau. (b) Variances ratio as function of the pump power. Solid line is a fit with the function in Eq.\ref{eq:uno} and the dashed line is a fit of the initial plateau. Indication of parametric induced noise squeezing is observed for pump power below 100 $\mu$W, in agreement with Ref. \cite{Guarrera2019}.}
\label{fig:Squeezing}
\end{figure}

\textit{Pump beam power dependence.} 
In the pump beam power range 1-4 $\mu$W the variances of two signal components X and Y do not significantly differ, Fig. \ref{fig:Spectrum}(a). For pump powers above 4 $\mu$W, instead, we start observing an asymmetry in the atomic noise distribution with the noise on the X (in-phase) quadrature being significantly greater than the noise on the Y (out-of-phase) quadrature. Figure \ref{fig:Spectrum}(b), for example, shows the spin noise spectrum recorded for a pump beam power of 21.5 $\mu$W where this asymmetry is clearly visible. Furthermore, for higher pump beam powers (i.e. above 50 $\mu$W with these lock-in amplifier settings), the SN spectrum shape of the Y component changes, with the generation of a dip in correspondence to the resonant frequency $\omega_{L}=\omega_0=\omega_M/2$, see Fig. \ref{fig:Spectrum}(c). We point out that the visibility of this dip depends on the time constant, and hence on the dynamical response of the lock-in amplifier, as we will further discuss in the next paragraph. The observation of asymmetry in the on-resonance noise distribution of conjugated spin variables is a signature of noise squeezing \cite{Guarrera2019}. In fact, we have compared these spectra with those acquired for the same pump powers close to a non-parametric frequency such as $\omega_L=\omega_M/4$ (and $\omega_{ref}=\omega_M/4$). In this case, the X and Y noise components are always similar, and no asymmetry nor the formation of a dip is observed. A quantitative comparison of the noise level, for the same pump power, has shown squeezing of the Y component with $\omega_L \simeq  \omega_M /2$. We note, however, that the values of the squeezed variances are always greater than those measured for a truly unpolarized spin state (i.e. in absence of the pump). Similar off-resonant quadrature spectra have been used for inferring (quantum) noise squeezing in mechanical systems \cite{Wollman2015,Pirkkalainen2015}. 

To isolate the source of the noise asymmetry from other possible non-linear effects in our noise spectra and help discriminating the dependence on $\omega_M$ and $\omega_{ref}$ in Fig. \ref{fig:Spectrum}(d) we analyze the ratio between the Y and X variances, $(\Delta F_Y)^2/(\Delta F_X)^2$ for different pump powers. With increasing pump powers two main identifiable features emerge in the spectra: a broad Lorentzian profile and a narrow near-resonance feature, black points in Fig. \ref{fig:Spectrum}(d). When the acquisition time is sufficiently long not to limit the spectral lines, their linewidths provide information about the coherence time of the noise process \cite{Crooker2004}. We thus analyse the FWHM of these two features, and we find that the FWHM of the broad feature grows with increasing pump power, inset (e) in Fig. \ref{fig:Spectrum}. In the same range of parameters, no such an effect of the pump power is detected in the Fourier analysis of the transient oscillation, i.e. varying only $\omega_{ref}$ \cite{Gartman2018}. This indicates that the Lorentzian profile mainly contains information about the noise dependence on the excitation frequency around $2\omega_{0}$. The measured functional dependence on the pump power is compatible with the linear scaling of the excitation frequency range with the amplitude of the parametric term for a detuned oscillator \cite{Landau,Charmichel1984}. Differently, the FWHM of the near-resonance feature reveals a more subtle dependence on the pump power, Fig. \ref{fig:Spectrum}(f). In particular, it shows a non-monotonic behaviour, and the FWHM decreases for increasing values of the pump in the interval $10-100$ $\mu$W. We note that this is the same range where the parametric effect has been observed in Ref.\cite{Guarrera2019}. Thus, interpreting this as a purely spectral feature (i.e. considering only the noise dependence on $\omega_{ref}$), allows us to draw some conclusion on the lifetime of the noise squeezing process, which increases as a function of the strength of the parametric term. This is clearly different from what is expected from power broadening due to the pump, and it is instead compatible with a resonant parametric effect in presence of damping \cite{Landau}.

Finally, even though in the timescale of these measurements the average signal $\langle F_X(t)\rangle=\langle F_Y(t) \rangle \approx 0$, we can deduce information about the system susceptibility to develop coherences oscillating at $\omega_{ref} \simeq \omega_L$ from the integrated noise, $A=\int (\Delta F)^2 d\omega$ \cite{Crooker2004}. We observe that $A$ grows linearly with the pump power above roughly $\sim 4$ $\mu$W, see Fig. \ref{fig:Squeezing}(a). The noise asymmetry on resonance $(\Delta F_Y)^2_0/(\Delta F_X)^2_0$, which also shows a threshold behaviour around $4$ $\mu$W, saturates at about $60$ $\mu$W, Fig. \ref{fig:Squeezing}(b). For the case of a parametric amplification via modulation of the relaxation rate of a classical oscillator we expect \cite{Briant2003}:
\begin{equation}
\frac{(\Delta F_Y)^2}{(\Delta F_X)^2}=\frac{1-g}{1+g}
\label{eq:uno}
\end{equation}
with $g$ the parametric gain. According to the model in Ref.\cite{Guarrera2019} the power broadening due to the amplitude modulated pump beam is responsible for a parametric contribution to the spin coherence relaxation rate, i.e.
\begin{equation}
g  \propto L_R WW^{\dagger} \rho \propto I_P
\label{eq:due}
\end{equation}
where  $W=-E_0(t) \Pi \textbf{d} \cdot \textbf{e}^* \Pi_e$ ($W^{\dagger}=E_0(t) \Pi_e \textbf{d} \cdot \textbf{e} \Pi)$ with $E_0(t)$, $\textbf{e}$, and $I_P=\vert E_0(t) \vert^2$ the amplitude, polarization versor, and intensity of the laser field respectively, and $\textbf{d}$ the induced atomic dipole moment. $\Pi$ and $\Pi_{e}$ are the projectors on the ground and excited state (labelled with the letter $e$) which are connected by the pump, $L_R$ depends on the natural linewidth of the transition, and $\rho$ is the system density matrix. A fit combining equations (\ref{eq:uno}) and (\ref{eq:due}) provides good agreement with our data in the range $4-50$ $\mu$W. This indicates that the noise asymmetry does not grow unbounded with the coherences generated by the pump, but the parametric effect soon saturates. 
\begin{figure}[h!]
\includegraphics[width=\columnwidth]{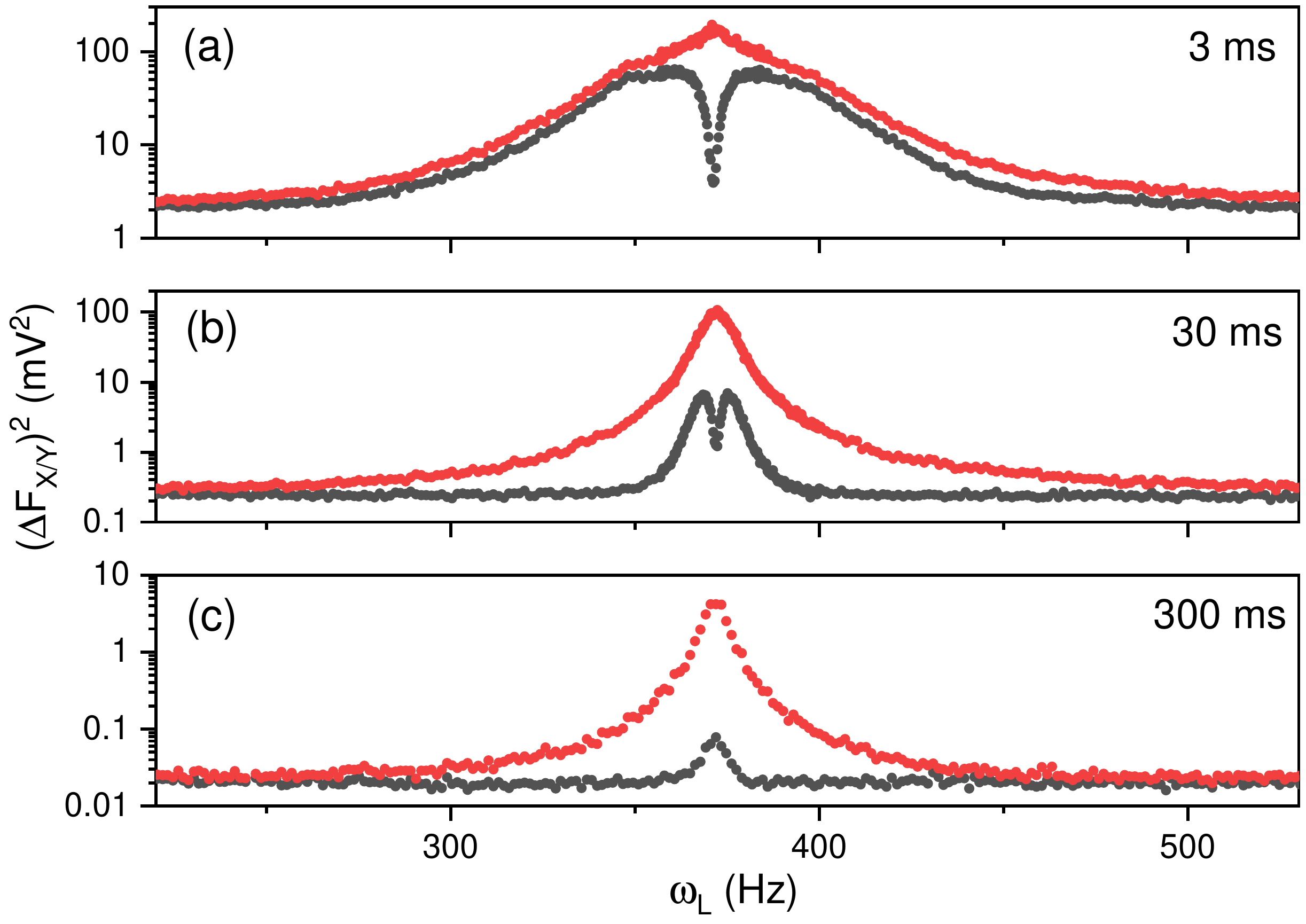}
\includegraphics[width=0.4\textwidth]{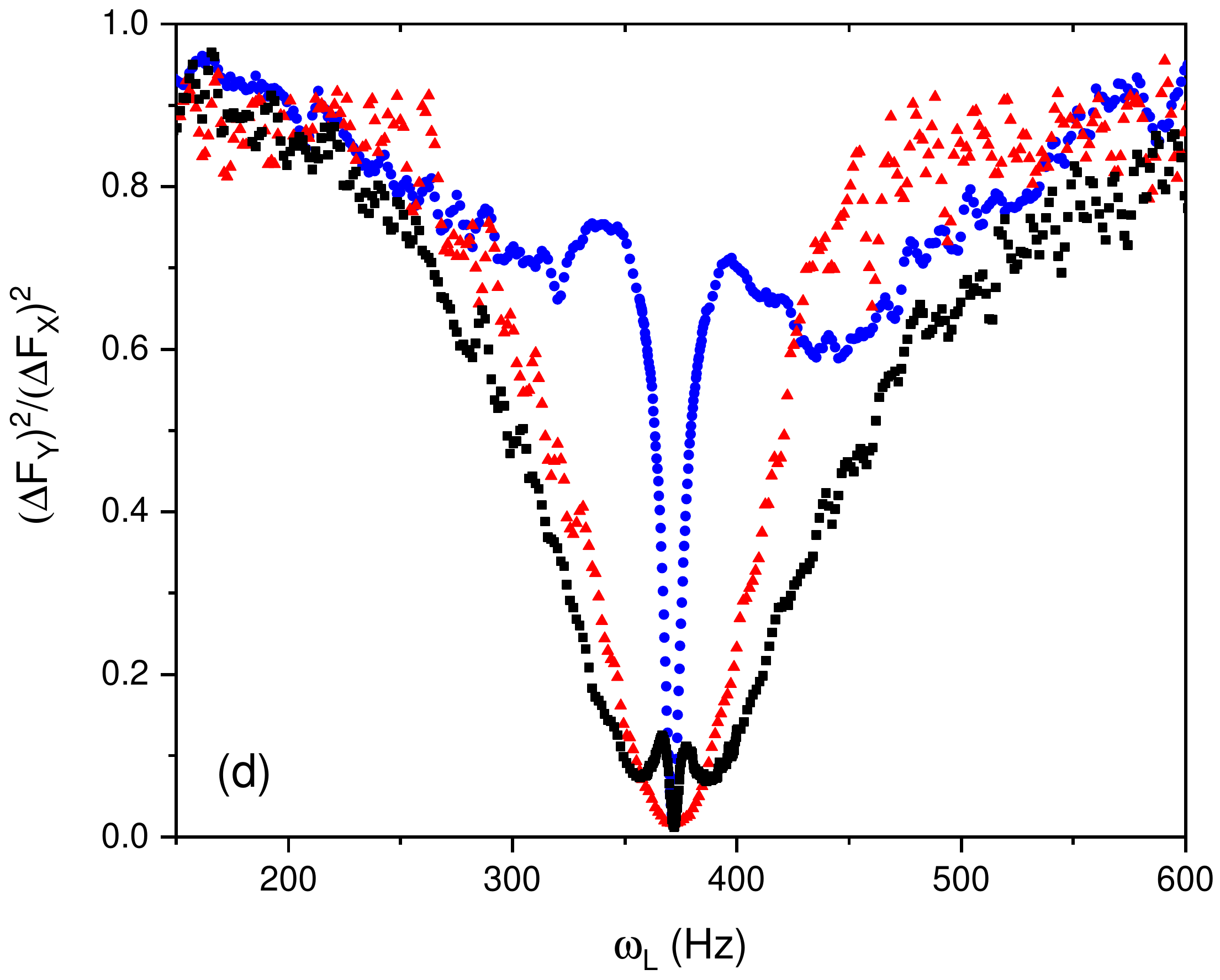}
\caption{ (color online) Dependence of the signal variance on Larmor frequency recorded with the lock-in time constant (a) 3, (b) 30, and (c) 300 ms (lock-in referenced to $\omega_{ref}=370$ Hz). (d) Dependence of the ratio of the two noise quadratures $(\Delta F_Y)^2/(\Delta F_X)^2$ on the lock-in time constant, with $\tau=3$ ms (blue dots), $30$ ms (black squares), and $300$ ms (red triangles). The measurements were made with pump power of 79 $\mu$W and probe power of 146 $\mu$W beam power.  }\label{fig:Time_constant}
\end{figure}

\textit{Temporal dependence.} 
In Figure \ref{fig:Time_constant} we show the noise spectra recorded with three lock-in time constants (a) $\tau=3$ ms, (b) $\tau=30$ ms and (c) $\tau=300$ ms, leaving the other parameters unvaried (i.e. pump and probe beam powers, and atomic density). The use of a relatively low time constants, such as $3$ ms, reduces the amount of time-averaging performed by the lock-in amplifier, and provides similar values of the X-Y variances outside the magnetic resonance, see Fig. \ref{fig:Time_constant} (a). Synchronization of the signal phase with the modulation frequency of the optical excitation and lock-in reference takes place in vicinity of the resonance, where the narrow structure appears in the spin noise spectra (Fig. \ref{fig:Time_constant}(a,b)). For a long lock-in time constant we observe strong out-of-resonance averaging of the noise fluctuations in the spectrum, Fig. \ref{fig:Time_constant} (c), while the spectrum immediately around the resonance appears more robust. An analysis of the ratio $(\Delta F_Y)^2/(\Delta F_X)^2$ shows a different time dependence of the two main components. In particular, the Lorentzian profile emerges from longer time averaging, as a narrower band of the atomic noise is filtered by the lock-in amplifier. Instead, the near-resonance feature becomes progressively less visible, as increasing the lock-in time constant corresponds to binning the signal on a timescale comparable ($\tau=30$ ms) or larger ($\tau=300$ ms) than the lifetime of the noise process.

\begin{figure}[h!]
\includegraphics[width=\columnwidth]{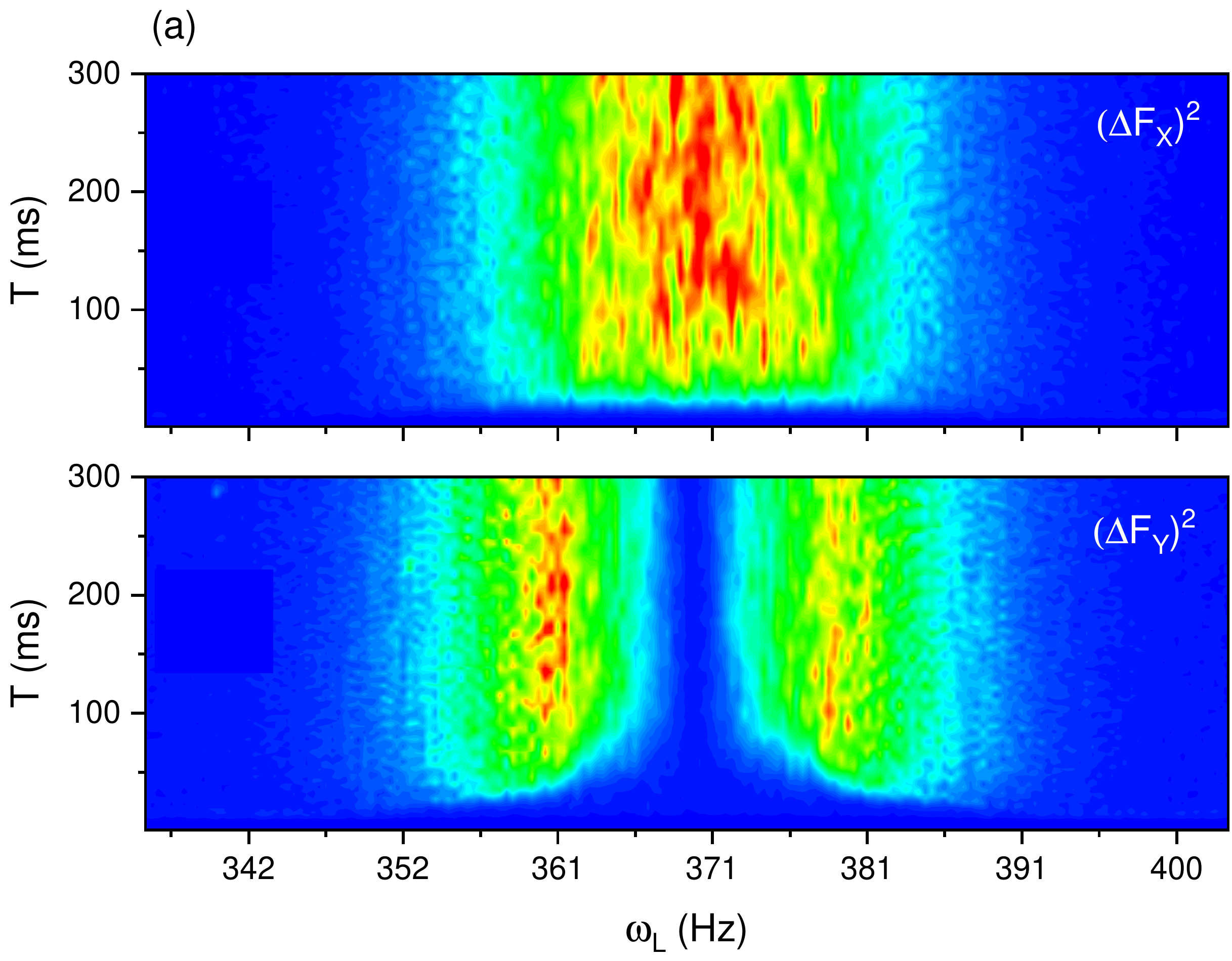}
\includegraphics[width=\columnwidth]{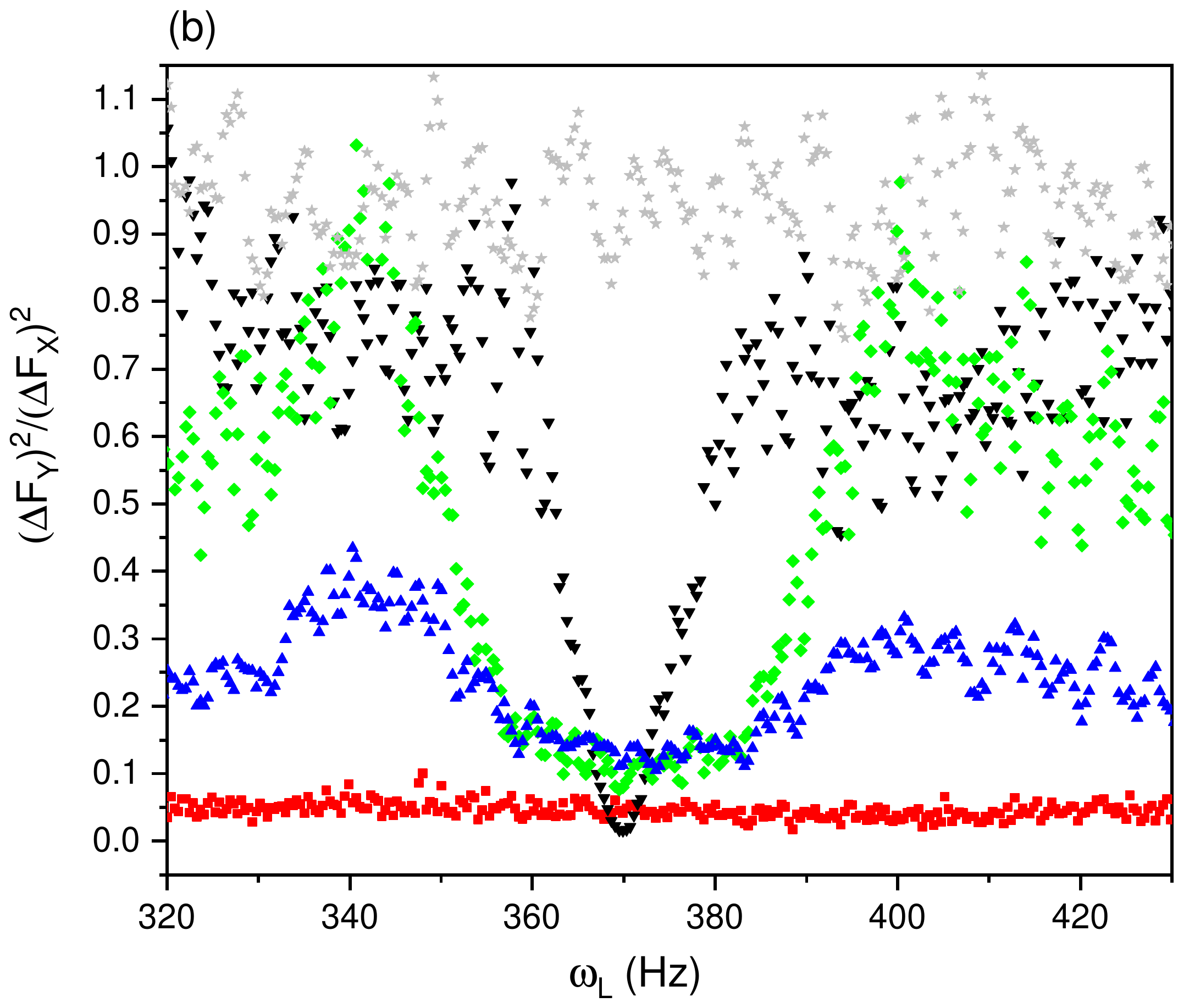}
\caption{ (color online) (a) Dependence of the noise spectra on the total acquisition time, recorded with lock-in time constant of $3$ ms (lock-in referenced to $\omega_{ref}=370$ Hz), and pump power of 79 $\mu$W. Note that the color scales are different for the two quadratures. (b) Spectra of the two noise components' ratio $(\Delta F_Y)^2/(\Delta F_X)^2$ for different acquisition times, starting from the beginning of the pumping process: $1$ ms (grey stars), $5$ ms (red squares), $10$ ms (blue triangles), $15$ ms (green diamonds), and $100$ ms (black upside-down triangle).  }
\label{fig:Time}
\end{figure}

To get further insight into the dynamics of the noise generation, we analyse the evolution of the noise spectra from the starting of the pumping process Fig. \ref{fig:Time}(a). The measured noise for the two quadratures mainly varies within the first $\sim 100$ ms and then does not change up to our maximum observation time T of several seconds. We note that in these measurements the amplitude of the signal at $\omega_L$ decays on a comparable timescale of $T_2=(94\pm 1)$ ms, which mainly depends on the rate of the spin-exchange collisions. The ratio between the two quadrature variances reveals an asymmetric noise distribution already after few ms time, Fig. \ref{fig:Time}(b). The near-resonance feature grows for longer observation times, while its FWHM decreases, up to roughly $T=100$ ms. This shows that the process generating the asymmetric noise distribution, and hence the noise squeezing, is limited by the acquisition time T up to roughly $100$ ms which is compatible with the spectrally derived lifetime of $1/$FWHM$=(86 \pm 2)$ ms and the dependencies on the lock-in time constant, discussed above. 

\textit{Conclusions.} 
The characterization of the noise spectrum that we have carried out confirms that the observed noise squeezing is due to the parametric excitation generated by the modulation of the pump beam power and it is mainly limited by the damping due to the finite lifetime of the atomic coherences.
Thermal atomic vapours create a platform for a wide range of precise atomic sensors with performances often limited by atomic projection noise \cite{Kitching2011}. 
Understanding of the processes which contribute to the signal/noise generation and their dynamics, assists in improving these devices \cite{Wasilewski2010}. The analysis of the noise spectra of out-of-equilibrium atomic ensembles allows not only to quantify the amount of noise squeezing, but it also provides insight into the processes that contribute to the generation of the noise distribution. We have verified this on a parametrically driven system, which extends the current application of SNS, and promotes SNS as a desirable tool for the investigation of spin-squeezing in non-linear, out-of-equilibrium systems.  

\color{black}
\begin{acknowledgements}
The work was funded by the UK Department for Business, Energy and Industrial Strategy as part of the National Measurement System Programme. V.G. was supported by the EPSRC (Grant No. EP/S000992/1). We would like to thank R. Hendricks and G. Barontini for critical reading of the manuscript. G.B. thanks V. Biancalana for fruitful discussions.
\end{acknowledgements}


\begin{thebibliography}{10}
\bibitem{note1} A small oscillation at $\omega_L$ has been recently observed in steady-state due to replenishing of the decaying coherence by off-resonant pumping.
\end{thebibliography}
\end{document}